\documentclass{emulateapj}
\usepackage{apjfonts}
\usepackage{natbib}

\def\Mo{{\rm M_\odot}}

\begin{document}

\shortauthors{CALLEGARI ET AL.}
\shorttitle{MBH GROWTH IN MINOR MERGERS OF DISK GALAXIES} 

%-----------------------------------------------------------------------------
\title{Growing Massive Black Hole Pairs in Minor Mergers of
  Disk Galaxies}
%-----------------------------------------------------------------------------

\author{Simone Callegari\altaffilmark{1}, Stelios
  Kazantzidis\altaffilmark{2}, Lucio Mayer\altaffilmark{1,3},
  Monica Colpi\altaffilmark{4},\\Jillian M. Bellovary\altaffilmark{5},
  Thomas Quinn\altaffilmark{6}, and James Wadsley\altaffilmark{7}}

\begin{abstract}
  We perform a suite of high-resolution smoothed particle
  hydrodynamics simulations to investigate the orbital decay and mass
  evolution of massive black hole (MBH) pairs down to scales of
  $\sim30$~pc during minor mergers of disk galaxies. Our simulation
  set includes star formation and accretion onto the MBHs, as well as
  feedback from both processes. We consider 1:10 merger events
  starting at $z \sim 3$, with MBH masses in the sensitivity window of
  the Laser Interferometer Space Antenna, and we follow the coupling
  between the merger dynamics and the evolution of the MBH mass ratio
  until the satellite galaxy is tidally disrupted. While the more
  massive MBH accretes in most cases as if the galaxy were in
  isolation, the satellite MBH may undergo distinct episodes of
  enhanced accretion, owing to strong tidal torques acting on its host
  galaxy and to orbital circularization inside the disk of the primary
  galaxy. As a consequence, the initial 1:10 mass ratio of the MBHs
  changes by the time the satellite is disrupted. Depending on the
  initial fraction of cold gas in the galactic disks and the geometry
  of the encounter, the mass ratios of the MBH pairs at the time of
  satellite disruption can stay unchanged or become as large as
  1:2. Remarkably, the efficiency of MBH orbital decay correlates with
  the final mass ratio of the pair itself: MBH pairs that increase
  significantly their mass ratio are also expected to inspiral more
  promptly down to nuclear-scale separations. These findings indicate
  that the mass ratios of MBH pairs in galactic nuclei do not
  necessarily trace the mass ratios of their merging host galaxies,
  but are determined by the complex interplay between gas accretion
  and merger dynamics.
\end{abstract} 


\altaffiltext{1}{Institute for Theoretical Physics, University of
  Z\"urich, Winterthurerstrasse 190, CH-9057 Z\"urich, Switzerland;
  callegar@physik.uzh.ch}  
\altaffiltext{2}{Center for
  Cosmology and Astro-Particle Physics; and Department of Physics; and
  Department of Astronomy, The Ohio State University, 191 West
  Woodruff Avenue, Columbus, OH 43210 USA.}  
\altaffiltext{3}{Institut
  f\"ur Astronomie, ETH Z\"urich-H\"onggerberg, Wolfgang-Pauli-Strasse
  16, CH-8093 Z\"urich, Switzerland.}  
\altaffiltext{4}{Dipartimento
  di Fisica G. Occhialini, Universit\`a di Milano Bicocca, Piazza
  della Scienza 3. I-20126 Milano, Italy.}
\altaffiltext{5}{Department of Astronomy, University of Michigan, 830 Dennison Bldg., 500 Church Street, Ann Arbor, MI 48109-1042, USA}
\altaffiltext{6}{Department of Astronomy, University of Washington,
  Box 351580, Seattle, WA 98195, USA.}  
\altaffiltext{7}{Department of
  Physics and Astronomy, McMaster University, 1280 Main Street West,
  Hamilton, ON L8S 4M1, Canada.}

\section{Introduction}
\label{introduction}

The ubiquity of massive black holes (MBHs) at the centers of galactic
spheroids \citep{richstone98} together with the ``bottom--up'' nature
of galaxy formation in the currently favored $\Lambda$CDM cosmology
\citep[e.g.,][]{whiterees78} suggest that MBH pairs at sub-kpc scales
should form in galactic nuclei during the hierarchical assembly of
structures (\citealt{begelman80}; see the recent review by
\citealt{colpi09b}).  Such pairs may undergo further orbital decay,
become Keplerian binaries, and eventually coalesce via the emission of
gravitational waves \citep{haehnelt94}, which will be one of the main
targets of the next generation of gravitational wave detectors such as
the Laser Interferometer Space Antenna (LISA)
\citep{vecchio04}. Moreover, observations indicate that the masses of
MBHs correlate with properties of their host spheroids, including the
luminosity, mass, and velocity dispersion
\citep[e.g.,][]{magorrian98,ferrarese00,gebhardt00}. Such relations
suggest the existence of fundamental physical mechanisms that link
MBH assembly and galaxy formation, and may connect the properties of
galaxy mergers with the resulting MBH binaries.

Given all these facts, the study of the dynamics of MBHs during galaxy
mergers becomes especially important as a means to connect the
cosmological assembly of galaxies with that of MBH pairs and
binaries. Many numerical studies have focused on the connection
between galaxy mergers and MBH growth, specifically in relation to the
MBH final mass, scaling relations and phenomenology of quasars
\citep[e.g.,][]{dimatteo05,hopkins05,boylan07,younger08,johansson08}. However,
much less attention has been devoted to investigating the orbital
decay and evolution of both MBHs {\it during} mergers, particularly in
the unequal-mass regime which comprises the vast majority of such
events.  \citet{kazantzidis05} and \citet{callegari09} (hereafter C09)
showed that the formation of unequal-mass MBH close pairs at sub-kpc
scales is sensitive to the details of the gas dynamics during the
merger process. In particular, they found that, due to the combination
of gas dynamics and star formation, a pair of MBHs can form
efficiently in 1:10 minor mergers at scales below 100~pc, provided
that galaxies are relatively gas-rich and that the mergers occur at
relatively high redshift ($z\sim3$) when halo densities are higher and
dynamical friction timescales correspondingly shorter, according to
the \citet{chandra43} standard formulation. However, these papers did
not model the accretion onto the MBHs, the evolution of their mass
ratio and its dependence on the merger dynamics. Such investigations
are important, as the mass ratio of MBH pairs is a fundamental
parameter which drives their evolution on nuclear scales
\citep[e.g.,][]{dotti07,lodato09}.

In this paper, we explore in detail for the first time the formation
and mass evolution of close MBH pairs down to scales of $\sim30$~pc
using controlled smoothed particle hydrodynamics (SPH) simulations of
minor mergers between disk galaxies. The merging systems have a mass
ratio of $1:10$ and our simulation suite includes the effects of star
formation and accretion onto the MBHs, as well as feedback from both
processes. We model merger events associated with MBH pairs whose
gravitational wave emission should be in the sensitivity window of
LISA: they have masses $\sim 10^5~\Mo$ around the epoch of a predicted
peak of MBH pair formation in mergers happening at $z \sim 3$
\citep{volonteri03,sesa05}.

\section{Numerical Simulations}
\label{s:sims}

Our reference galaxy model is a $z=3$ progenitor of a Milky Way-type
disk galaxy, with a virial velocity $V_{\rm
  vir}=145$~km~s$^{-1}$. Such a progenitor is constructed assuming a
constant $V_{\rm vir}$ between redshift 0 and 3 \citep{li07}, and
rescaling the mass and size of a $z=0$ model by the ratio
$\left[H(z=3)/H_0\right]=0.224$ of the Hubble constant at $z=3$ over
its present-day value \citep{mmw98} for a $\Lambda$CDM concordance
cosmology ($H_0=70$~km~s$^{-1}$~Mpc$^{-1}$, $\Omega_{\rm m}=0.3$,
$\Omega_\Lambda=0.7$). The model consists of three components: a dark
matter halo, a stellar and gaseous disk, and a stellar bulge. The halo
is a spherical \citet{nfw96} model with a virial mass of $M_{\rm
  vir}=2.3\times10^{11}\Mo$, a spin parameter $\lambda=0.04$
\citep{vitvit02}, and a concentration $c=3$ consistent with the
relations found by \citet{bullock01}. The halo has been adiabatically
contracted to respond to the growth of the disk and bulge
\citep{blumenthal86}. The baryonic disk follows an exponential
profile, with a total mass $M_d=0.04 M_{\rm vir}$, a radial
scale-length $R_d = 1.1$~kpc determined according to \citet{mmw98}, a
scale-height of $z_d =0.1 R_d$, and a mass fraction in gas denoted by
$f_{\rm g}$. The bulge is a spherical \citet{hernquist90} model with a
mass $M_b = 0.008 M_{\rm vir}$, yielding a ratio of bulge-to-total
stellar mass $B/T\sim0.2$ (depending on $f_{\rm g}$), and a scale
radius $a_b = 0.2 R_d$. The satellite galaxies are constructed
following the same relations between structural parameters and a mass
in each component scaled down by $q_{\rm gal}=0.1$, with a disk
scale-length of 300~pc.

The $N$-body realizations consist of $10^6$ particles in the dark
matter halo and $2\times10^5$ stellar particles in each galaxy. In
addition, each galaxy is initialized with $10^5$ gas particles in the
disk, except for one case (see Table~\ref{t:summary}). We adopted a
force resolution of $45$~pc for both the dark matter and baryonic
particles of the larger galaxy, while for the satellite galaxy we used
$20$~pc.

A particle representing the MBH was added at the center of each galaxy
with a mass according to the updated $M_{\rm BH} - M_{\rm bulge}$
relation of \citet{haringrix04}. Our choice for the galaxy masses in
conjunction with the assumption that the MBHs follow the $M_{\rm BH} -
M_{\rm bulge}$ relation result in MBH masses of $6\times10^4~\Mo$ and
$6\times10^5~\Mo$ for the satellite and primary MBH respectively, with
an initial MBH mass ratio $q=q_{\rm gal}$.  With these choices, we
target the typical masses and cosmic epoch of coalescing MBHs that
should be detectable by LISA \citep{volonteri03,sesa05}.

\begin{deluxetable}{cccc}
  \tablecaption{Summary of Merger Simulations \label{t:summary}}
  \tablehead{ \colhead{$f_{\rm g}$ primary} & \colhead{$f_{\rm g}$
      satellite} & \colhead{$\theta$~\tablenotemark{b}} & \colhead{
      notes } } \startdata
  30\% & 30\% & 0$^\circ$ & reference run \\
  30\% & 50\% & 0$^\circ$ & $1.7\times10^5$ SPH particles in the satellite \tablenotemark{a} \\
  10\% & 10\% & 0$^\circ$ & - \\
  30\% & 30\% & 20$^{\rm o}$ & - \\
  30\% & 30\% & 45$^{\rm o}$ & - \\
  30\% & 30\% & 0$^\circ$ & small pericenter (0.8~kpc)
\enddata
\small{ \tablenotetext{a}{This number has been chosen in order to have
    the same ratio of gas particle mass to MBH mass in both primary
    and satellite galaxies, as happens for all other simulations here
    discussed.}  \tablenotetext{b}{The orbital plane and satellite
    disk are inclined by an angle $\theta$ with respect to the disk of
    the primary galaxy.}  }
\end{deluxetable}

We choose merger orbital parameters that are common for merging halos
in cosmological simulations \citep{benson05}. All simulations start
with the baricenters of the two galaxies at a distance equal to the
sum of their virial radii. Their orbit is initially parabolic, with
pericentric distance $R_{\rm p}=8$~kpc, equal to 20\% of the primary
galaxy's virial radius, except for one case with a 10 times smaller
pericenter. We also vary gas fractions and orbital inclinations. A
summary of the simulations here presented can be found in
Table~\ref{t:summary}.\footnote{ We note that only prograde mergers
  have been considered in this work. In fact, retrograde encounters
  are characterized by a much slower orbital decay and, particularly
  in the low mass ratio regime, produce naked ``wandering'' MBHs as
  the satellite galaxy is disrupted at distances $\sim10$~kpc. From
  such distances the dynamical friction timescale for producing a MBH
  pair at $\sim100$~pc are longer than a Hubble time.}  All
simulations were performed with GASOLINE, a multi-stepping, parallel
$N$-body/SPH code \citep{gasoline}. We include atomic cooling for a
primordial mixture of hydrogen and helium, and the star formation
algorithm is based on the local Schmidt-Kennicutt law
\citep{katz92}. Feedback from supernovae is treated using the
blast-wave model described in \citet{stinson06}. Accretion onto the
MBHs is also modelled with a sub-grid recipe
\citep[e.g.,][]{springel05}: the accretion rate $\dot M_{\rm BH}$ is
estimated from the density $\rho_g$, sound speed $c_s$ and relative
velocity $V$ of the gas within a smoothing length from the MBH, via a
Bondi-Hoyle-Lyttleton type formula, $\dot M_{\rm BH} = 4 \pi G^2
M_{\rm BH}^2 \rho_g (V^2 + c_s^2)^{-3/2}$ \citep{bondi52}. Of this
mass-energy input, a fraction $\epsilon_r=0.1$ is assumed to be
radiated away, while a fraction $(1 - \epsilon_r)=0.9$ is added to the
mass of the MBH from its neighboring gas particles.  Lastly, a
fraction $\epsilon_{\rm fb}=0.005$ of the radiated luminosity couples
to the surrounding gas as a heating source.  This feedback efficiency
is tuned so that a number of constraints related to our initial galaxy
models can be satisfied (see Section~\ref{ss:fg03}).  In principle,
the feedback efficiency depends both on the employed sub-grid model
for the interstellar medium and the numerical resolution. For all
these reasons, our specific choice differs from those used in a number
of earlier studies \citep[e.g.,][]{springel05}. Since the simulations
are halted at satellite disruption, the orbital decay of the satellite
MBH is determined by the well-resolved drag forces acting on its host
galaxy; therefore, we do not include a sub-resolution Bondi drag term,
as employed recently by, e.g., \citet{younger08}. Analogously, we do
not employ any other numerical recipe that forces MBH orbital decay,
because our aim is to study the growth of MBHs \emph{in relation to
  the efficiency of close MBH pair formation}. In particular, the
black holes were not repositioned at the local minimum of the
gravitational potential at each time-step
\citep[e.g.,][]{johansson08}.

\section{Results}
\label{s:results}

In Section~\ref{ss:fg03} our reference simulation is discussed, where
the gas fraction in both disks of the primary and satellite galaxy is
$f_{\rm g}=0.3$. Sections~\ref{ss:gasfrac} and \ref{ss:orbits} compare
this reference case with other merger simulations of different gas
fraction and orbital parameters, respectively, as summarized in
Table~\ref{t:summary}.

\subsection{Reference Simulation: Coplanar Merger with $f_{\rm g}=0.3$}
\label{ss:fg03}

\begin{figure}
\epsscale{1.2}
\plotone{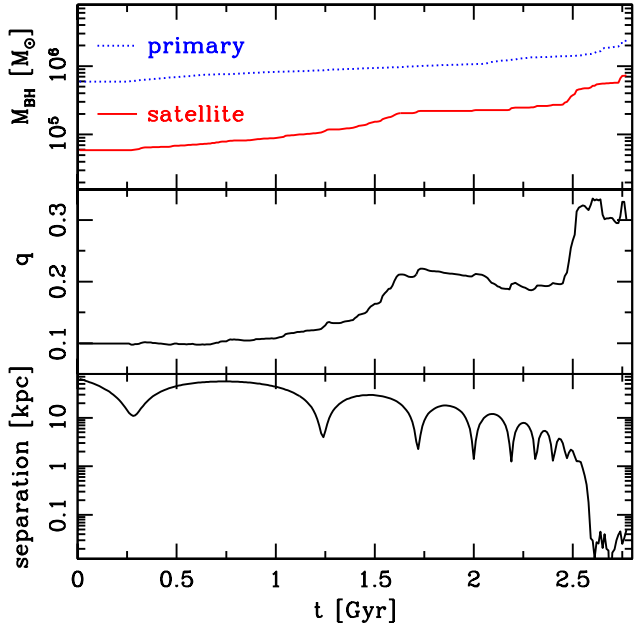}
\caption{Coplanar merger with a disk gas fraction of $f_{\rm g}=0.3$ in both
galaxies.  {\it Upper panel}: Evolution of the mass of the MBH,
$M_{\rm BH}$, in the primary (dotted line) and satellite galaxy (solid
line) as a function of time. {\it Middle panel}: Evolution of the mass
ratio $q$ of the two MBHs as a function of time. {\it Lower panel}:
Orbital decay of the two MBHs as a function of time.
\label{fig:fg03}}
\end{figure}

In order to assess the effects of the merger on the mass growth of the
MBHs, we first performed simulations of the primary and satellite
galaxies evolved {\it in isolation}. For these tests, we chose the
$f_{\rm g}=0.3$ galaxy models of our reference simulation. In these
simulations, the mass ratio of the two MBHs in the isolated galaxies
does not deviate significantly or systematically from the initial
$q=0.1$ (the maximum fluctuation around this value is $\sim 10\%$)
over a period of more than $2$~Gyr. Therefore, we satisfy our working
hypothesis that, in equilibrium conditions, the $q$ of the black holes
corresponds to the galaxy mass ratio, following from our initial
choice of $M_{\rm BH} - M_{\rm bulge}$ and galaxy morphology. In
addition, these tests provide a measure of the MBH ``quiescent''
accretion: mass-doubling timescales of the MBHs evolved in isolation
are $\sim2$~Gyr, i.e. comparable to the typical duration of one of the
mergers presented below. This result indicates that any larger MBH
growth or large variation of $q$ during the merger does not stem from
secular evolution in the galaxy models or from numerical effects, but
rather should be attributed to the galaxy encounter itself.

\begin{figure}
\epsscale{1.2}
\plotone{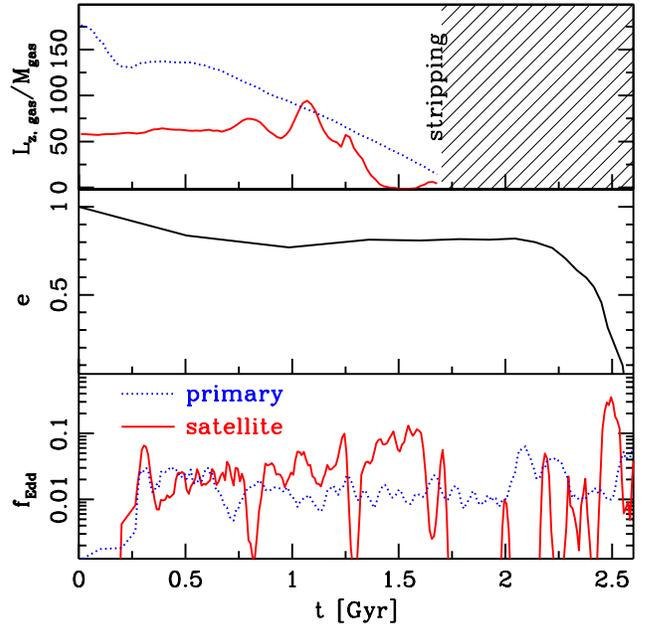}
\caption{Evolution of properties related to the coplanar merger with
  $f_{\rm g}=0.3$ in both galaxies.  {\it Upper panel}: Gas specific angular
  momentum in the direction of the disk rotation axis in the primary
  (dotted line) and companion galaxy (solid line) (see text for
  details). {\it Middle panel}: Orbital eccentricity of the satellite
  MBH inside the disk of the primary.  {\it Lower panel}: Eddington
  ratios, $f_{\rm Edd}$, of the mass accretion rates onto the primary
  (dotted line) and secondary MBH (solid line). \label{fig:ecc}}
\end{figure}

Figure~\ref{fig:fg03} presents the mass evolution of the two MBHs, the
evolution of their mass ratio, and their relative separation as a
function of time. By the end of the merger ($t\sim 2.6$~Gyr), owing to
dynamical friction, the two MBHs have formed a close pair in the
nucleus of the remnant at a separation comparable to the adopted force
resolution. This finding confirms our previous results (C09) and
suggests that gas accretion onto the MBHs and associated feedback is
not critical for pair formation in this case. The primary MBH grows
quiescently throughout most of the merger, while the secondary
increases its mass tenfold by the time the pair forms. The
corresponding increase in the mass ratio of the two MBHs, $q$, occurs
in two distinct episodes, which are elucidated in
Figure~\ref{fig:ecc}.

The upper panel of Figure~\ref{fig:ecc} shows the evolution of the gas
specific angular momentum in the direction of the disk rotation axis,
$L_{z, {\rm gas}}/M_{\rm gas}$, as a function of time for both
galaxies. For this calculation, the gas particles that contribute to
the baryonic mass within the central 5 softening lengths of each
galaxy right before the third pericentric passage ($t=1.7$~Gyr) were
traced back in time. This figure shows that most of the gas in the
nuclear region of the satellite at this stage has lost its specific
angular momentum on a relatively short timescale. The reason for the
angular momentum loss is the strong tidal torques that occur near the
second pericentric passage ($t=1.2$~Gyr) and are induced by the
gravitational interaction with the primary galaxy.  As a result, a
$\sim 0.5$~Gyr long accretion episode with a corresponding increase of
$q$ is observed.  As shown in the lower panel of the same Figure, the
accretion rate $\dot M$ onto the satellite MBH is $10\%$ of the
Eddington limit $\dot M_{\rm Edd}$ \citep{eddington16} during this
phase.

On the other hand, the gas that is funnelled near the center of the
primary galaxy has been experiencing a nearly steady loss of specific
angular momentum over a very long timescale
(Figure~\ref{fig:ecc}). This indicates that angular momentum loss in
the case of the primary galaxy is not caused by tidal torques arising
from the interaction with the satellite, which would occur at
pericentric passages and become stronger as the merger progresses.
Rather, it is induced by secular evolution (i.e. spiral arms) which
redistributes angular momentum throughout the disk. In this context,
the effect of initial transient spiral arms is evident during the
first $\sim 200$~Myr. This agrees with the fact that mass growth of
the primary MBH is essentially unchanged between the merger and the
evolution in isolation.

\begin{figure*}
\epsscale{1.2}
\plotone{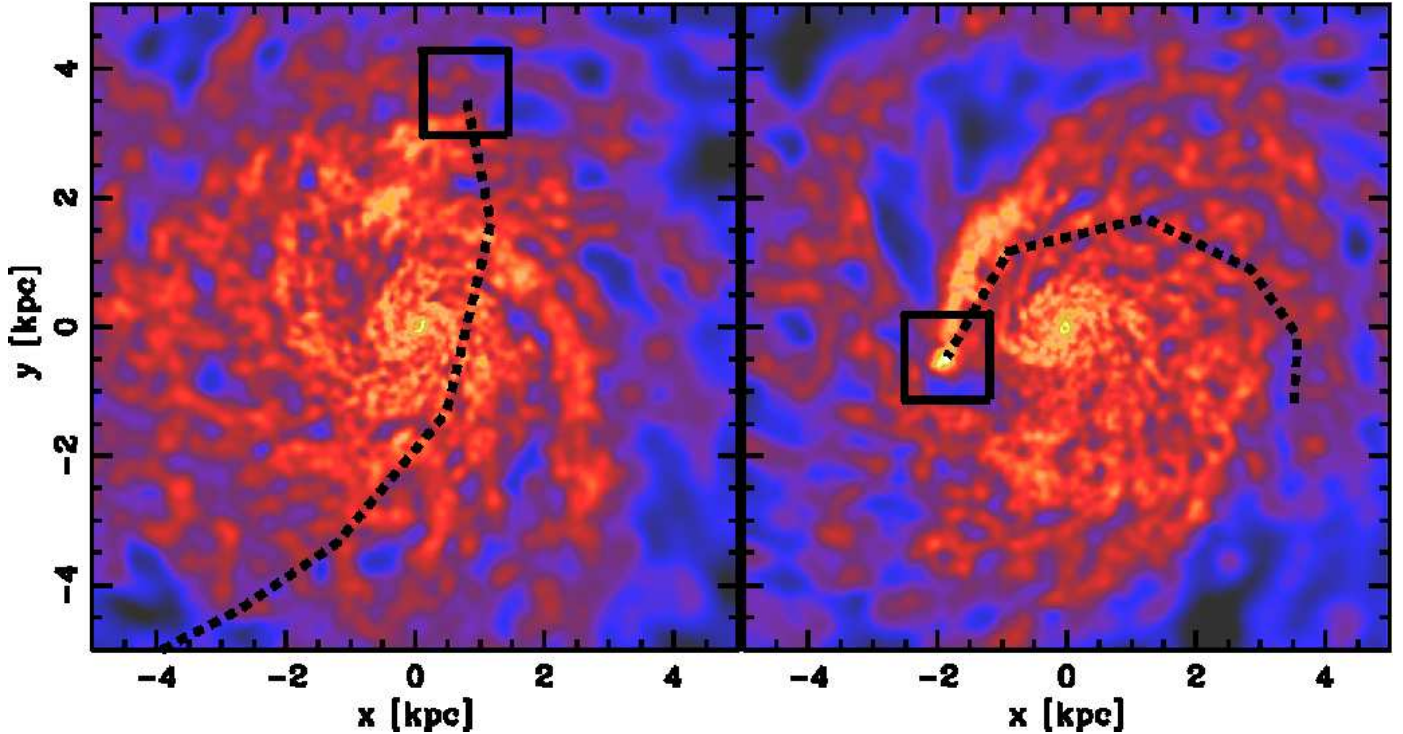}
\caption{Gas density maps at $t=2.33$ (left panel) and $t=2.48$~Gyr
  (right panel), which correspond to just before and after the orbit
  of the secondary MBH circularizes inside the disk of the
  primary. The time span between these two snapshots corresponds
  roughly to two orbits of the satellite. The maps show the inner
  $5$~kpc of the primary. Density is projected onto the $x-y$ plane
  and is color-coded on a logarithmic scale with brighter colors
  corresponding to regions of higher gas density. The black dashed
  line marks the trajectory of the satellite MBH and the square
  indicates the region around it. The satellite does not appear in the
  left panel, as ram pressure has stripped its entire gas content. On
  the contrary, the position of the satellite in the right panel is
  evidently traced by the overdensity and wake excited in the primary
  disk. \label{fig:gasmap}}
\end{figure*}

Around the third pericentric passage, ram pressure exerted by the
interstellar medium of the primary galaxy strips all the gas from the
satellite down to our force resolution. This is in agreement with
analytic estimates based on the study by \citet{marcolini03}, as
discussed also in C09. As a result, the satellite is now devoid of gas
(as shown in left panel of Figure~\ref{fig:gasmap}), and accretion
onto the smaller MBH is suddenly halted. A period of slowly decreasing
$q$ follows. In fact, during this phase, the more massive MBH
continues to accrete gas from its host, experiencing an increase in
its Eddington ratio $f_{\rm Edd}\equiv\dot M_{\rm BH}/\dot M_{\rm
  Edd}$ as the satellite galaxy is now orbiting close enough to excite
gas inflows in the primary disk ($t>2$~Gyr).

The mass of the satellite MBH sharply increases again (with an
associated second increase in $q$) at kiloparsec-scale separations. At
this stage, the secondary MBH orbits inside the gaseous disk of the
primary. The mass increase coincides with a sudden drop in the orbital
eccentricity of the satellite MBH, as shown in the middle panel of
Figure~\ref{fig:ecc}. This drop in eccentricity is caused by dynamical
friction acting on the satellite along its prograde coplanar orbit in
the high-density region of the primary disk. Such orbit
circularization is analogous to that found for MBHs in circumnuclear
disks \citep{dotti09a}. Thus, the satellite MBH and its host stellar
cusp are moving with a low relative velocity with respect to the disk
of the primary. As a consequence, they are able to collect surrounding
gas with low angular momentum in the reference frame of the
satellite, creating an overdensity (right panel of
Figure~\ref{fig:gasmap}) from which material is efficiently accreted
by the satellite MBH up to a peak $f_{\rm Edd} \sim 0.3$. On the other
hand, accretion onto the primary MBH still relies on angular momentum
transport by instabilities in the disk. Such instabilities, triggered
by the sinking satellite, are stronger than at earlier times, but
still not able to sustain high $f_{\rm Edd}$.

Overall, by the time the satellite is disrupted reaching the nuclear
region of the merger remnant and the two MBHs form a close pair, the
combination of strong tidal torques and orbital circularization acting
on the companion galaxy has caused the MBH mass ratio to increase from
$1:10$ to $1:3$, bringing it into a regime of ``major'' mass ratio.

\subsection{The Role of Gas Fraction}
\label{ss:gasfrac}

\begin{figure}
\epsscale{1.2}
\plotone{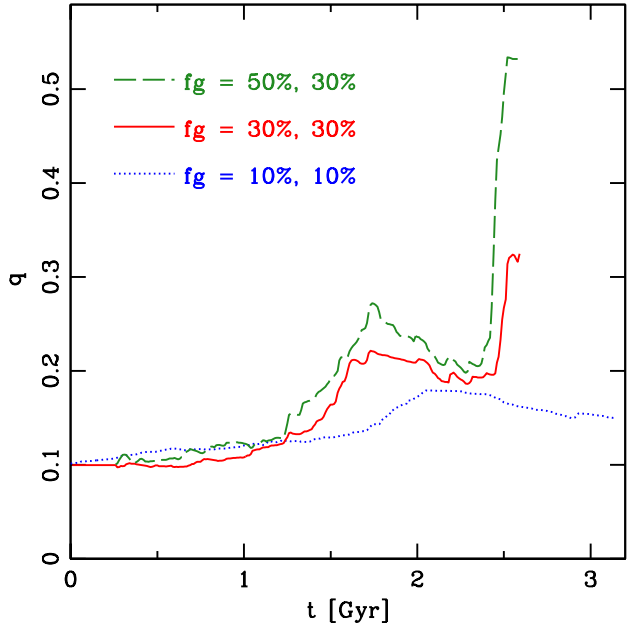}
\caption{Evolution of the mass ratio $q$ of the two MBHs as a function
  of time in coplanar mergers with different gas fractions $f_{\rm
    g}$. Dotted (blue) and solid (red) lines show results for the
  mergers with gas fractions $f_{\rm g}=0.1$ and $f_{\rm g}=0.3$,
  respectively, in both galaxies. The dashed (green) line corresponds
  to a merger where the initial gas fractions are $f_{\rm g}=0.3$ in
  the primary and $f_{\rm g}=0.5$ in the satellite
  galaxy. \label{fig:q}}
\end{figure}

Figure~\ref{fig:q} compares the evolution of the mass ratio of the
MBHs $q$ as a function of time in mergers with different disk gas
fractions, $f_{\rm g}$ (see Table\ref{t:summary}). All of these
mergers are on prograde, coplanar orbits. Again, we follow the
evolution of the interacting systems up to the time when the orbital
decay of the satellite galaxy is completed. This figure shows that the
first phase of increase in $q$, caused by dynamical destabilization of
the satellite, happens in all three cases: due to the fact that
inclination and orbit are fixed in these mergers, the torques acting
on the satellites have the same strength. Interestingly, the highest
value of $q$ reached in the first stage traces roughly the amount of
gas $\propto f_{\rm g}$ available for MBH fuelling. In all the mergers
considered here, ram pressure is effective in removing gas from the
satellite galaxy, once the two galaxy disks come into contact. When
this happens, accretion onto the secondary MBH is halted.

Figure~\ref{fig:q} shows that the initial $f_{\rm g}$ does cause
significant differences in the evolution of $q$ also during the second
stage of the mergers. Indeed, the second phase of strong accretion
onto the satellite MBH is absent in the merger with the smallest gas
fraction ($f_{\rm g}=0.1$). As discussed earlier, mass growth during
this phase becomes efficient only when the secondary MBH moves inside
the gas disk of the primary galaxy and its orbit
circularizes. Instead, in the case of $f_{\rm g}=0.1$ the satellite
does not sink below $\sim 400$~pc before being tidally disrupted,
because it has not experienced strong gas inflows steepening its
potential well during the first orbits (see C09 for a detailed
discussion). At such large distances, the MBH orbit is still mildly
eccentric, and the background density is not high enough to allow for
the formation of an overdensity that could trigger the second
accretion episode. As a consequence, the MBH mass ratio at the time of
satellite disruption is 1:6, close to its initial value, and the naked
satellite MBH will still take a few billion years to sink to the
center of the remnant via dynamical friction (C09).

We now focus on the merger where the initial gas fraction in the
primary and companion galaxies is equal to $f_{\rm g}=0.3$ and $f_{\rm g}=0.5$,
respectively.  Figure~\ref{fig:q} shows that this case is
characterized by a much larger final increase in $q$ compared to our
reference case, where the initial gas fraction was equal to $f_{\rm g}=0.3$
in both galaxies. Bearing in mind that the only difference between the
two initial conditions is the satellite $f_{\rm g}$, and that the satellite
gas has already been entirely stripped by ram pressure at this late
stage, this interesting result can be explained as an effect of the
different structure of the surviving satellite core and the different
mass growth of its MBH. In fact, a larger initial gas fraction allows
the satellite to build a denser stellar core via star formation in
response to tidal perturbations during the first two
orbits. Consequently, the nuclear region of the satellite is denser
and more massive. It is therefore more efficient at collecting gas
from the disk of the primary and it is subject to an enhanced
sinking. As a result, the orbit of the secondary MBH undergoes
circularization in a denser region of the primary disk, and is able to
accrete more gas. Moreover, by the time the secondary MBH enters the
disk of the primary, it is $\sim 60\%$ more massive compared to the
case of $f_{\rm g}=0.3$. This difference in mass naturally enhances its
accretion rate, which scales as $\propto M_{\rm BH}^2$. At the end of
this merger, the satellite MBH has sunk down to the center of the
remnant, where a close MBH pair is formed at scales $\sim30$~pc,
comparable to our force resolution. The mass ratio of the pair at this
point is the highest among all our simulations: $q=0.5$.

\subsection{The Role of Orbital Parameters}
\label{ss:orbits}

\begin{figure}
\epsscale{1.2}
\plotone{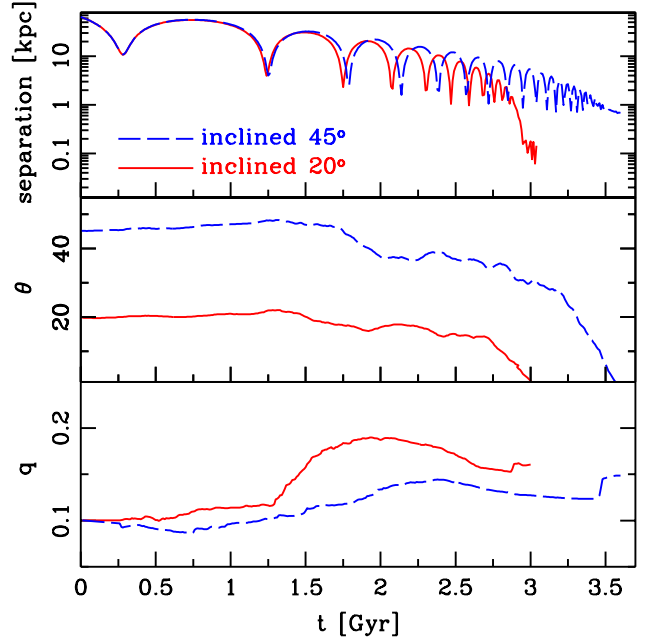}
\caption{The evolution of orbital parameters as a function of time is
  shown for the inclined runs: the continuous (red) line shows data
  for the merger with $20^\circ$ inclination, the dashed (blue) line for
  the one at $45^\circ$. {\it Upper panel}: MBH separation. As expected,
  sinking is more effective for more coplanar orbits. {\it Middle
    panel}: angle $\theta$ between the orbital plane of the two MBHs
  and the plane of the primary galaxy's disk. Orbit dragging onto the
  disk plane can be seen in both cases, once the satellite's orbit
  reaches the disk-dominated region of the primary galaxy. {\it Lower
    panel}: Mass ratio of the MBHs. \label{fig:inc}}
\end{figure}

As shown in Sections~\ref{ss:fg03} and \ref{ss:gasfrac}, the evolution
of the MBH mass ratio depends on processes (especially ram pressure
stripping of gas and orbit circularization) that may be sensitive to
the initial orbit of the merger itself. In this Section, we discuss
results from three simulations addressing the dependence on orbital
parameters. In all of them, the gas fraction in the two galaxies was
equal to that of the reference case, $f_{\rm g}=0.3$. 

Upper panel of Figure~\ref{fig:inc} shows the evolution of the MBH
orbital separation in two merger simulations where the orbital plane
and satellite disk were inclined with respect to the plane of the
primary disk, by $20^\circ$ in one case, and $45^\circ$ in the
other. Torques acting on the satellite during the early phases of the
merger are weaker for higher inclinations \citep{barneshernquist96},
and for this reason the increase in mass ratio $q$ during the first
three orbits is milder than in the coplanar case with the same gas
fraction (Figure~\ref{fig:inc}, lower panel). Moreover, a higher
inclination corresponds to a slower orbital decay; in fact, the
satellite spends only a small fraction of each orbit in the high
density, co-rotating baryonic region of the primary galaxy, where
dynamical friction is most efficient. The orbit is eventually dragged
down to the plane of the primary disk \citep{quinn86} (see middle
panel Figure~\ref{fig:inc}). However, since such drag takes a number
of orbits to cause significant alignment, the circularization effect
-- which can only happen when orbiting for most of the time inside the
primary disk -- is delayed compared to the coplanar case. For this
reason, the satellite undergoes a larger number of tidal shocks and is
eventually disrupted, precluding a second episode of substantial
accretion onto the satellite MBH. For low initial inclination
($\theta=20^\circ$), the drag is effective enough to bring the MBH
down to a separation of $\sim 120$~pc before the satellite is
disrupted. If the inclination is higher ($\theta=45^\circ$), the
satellite is disrupted leaving the MBH at $\sim 700$~pc distance,
where the dynamical friction timescale for sinking of a naked MBH is
of a few billion years.

\begin{figure}
\epsscale{1.2}
\plotone{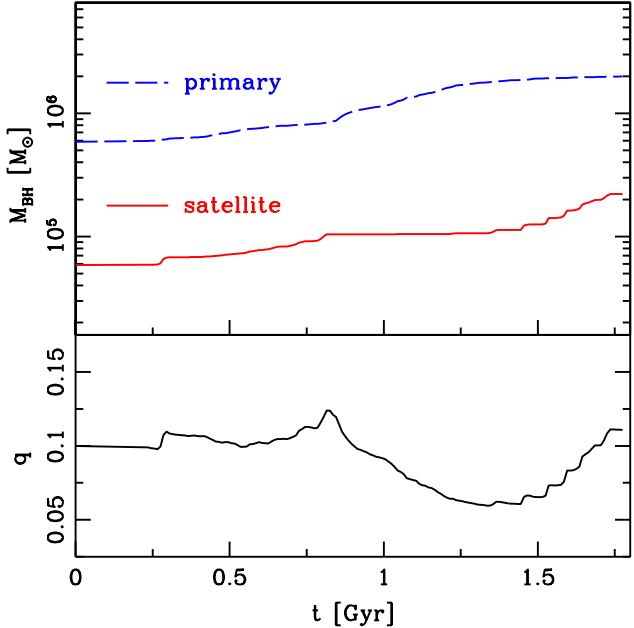}
\caption{MBH evolution in the $f_{\rm g}=0.3$ merger with $0^\circ$ inclination
  and small pericenter. {\it Upper panel}: mass accretion of the
  MBHs. The growth of the satellite MBH stops at second passage
  ($t=0.8$~Gyr), earlier than in the other cases, as ram pressure
  stripping of the satellite gas is more efficient at smaller
  separations. The primary MBH, on the other hand, starts a period of
  enhanced accretion after the second orbit, as the perturbations
  excited onto the primary galaxy are correspondingly stronger. {\it
    Lower panel}: evolution of the MBH mass ratio
  $q$. \label{fig:peri}}
\end{figure}

Finally, we turn to the merger starting on a parabolic orbit with very
small pericenter ($0.8$~kpc). The MBH masses and $q$ as a function of
time in this case are shown in Figure~\ref{fig:peri}. During the first
orbits, the satellite in this merger is subject to stronger tidal
torques than in the reference case, but MBH accretion is weaker. This
is because the ram pressure acting on the satellite's gas is $P\propto
\rho_{\rm ext} V^2$, where $\rho_{\rm ext}$ is the gas density in the
primary disk and $V$ is the relative velocity between the satellite
and its surroundings. Because of the small distance (corresponding to
a high $\rho_{\rm ext}$) and high $V$ at pericenters, ram pressure
stripping is effective since the very first passage: the satellite MBH
is starved early and is unable to accrete a substantial amount of
mass. For the same reason, MBH growth in the satellite is halted
completely already at second pericenter ($t\sim 1.8$~Gyr). On the
contrary, the primary MBH experiences enhanced accretion: in fact, the
close passages of the satellite excite strong instabilities in the
primary disk, funnelling gas towards its nucleus. For this reason, at
intermediate times during the merger, the mass ratio drops down to
$1:20$. Analogously to the reference run, a second phase of accretion
onto the satellite MBH occurs, in this case bringing $q$ back to its
starting value. However, MBH sinking does not proceed any further: as
a consequence of the strong tidal shocks experienced by the satellite
at pericentric passages, the satellite is disrupted leaving its MBH at
a distance of 1~kpc. At these separations, the estimated timescale for
orbital sinking of the naked MBH is longer than a Hubble time.

\section{Discussion and Conclusions}
\label{s:conclusions}

\begin{figure}
\epsscale{1.2}
\plotone{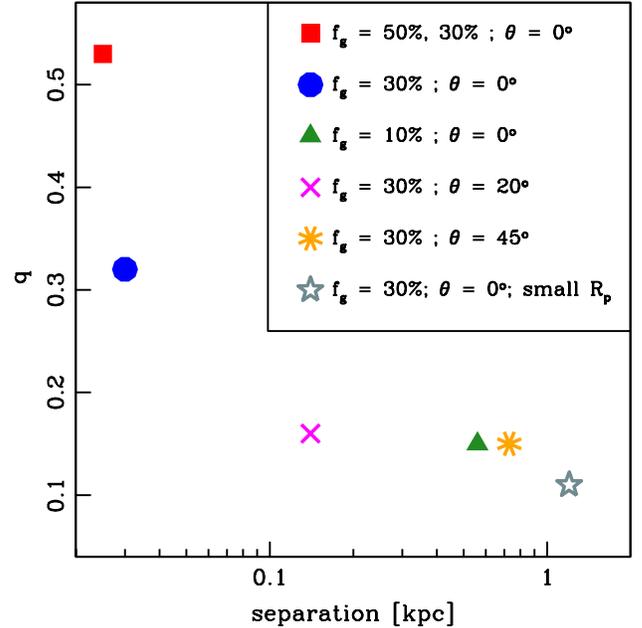}
\caption{The different symbols show the separation of the two MBHs and
  the corresponding MBH mass ratio $q$ at the time of satellite
  disruption for all the 1:10 mergers discussed here, labelled
  according to their initial gas fractions $f_{\rm g}$, orbital
  inclination $\theta$ and initial pericenter $R_{\rm
    p}$. \label{fig:final}}
\end{figure}

The outcomes of the simulations presented in this Paper are summarized
in Figure~\ref{fig:final}, where the mass ratio $q$ of the MBHs is
plotted against their separation, at the time of satellite
disruption. As discussed in the previous Sections, the physical
processes that facilitate the MBH orbital decay correspond to a
stronger accumulation of gaseous mass in the central region of the
satellite, compared to the isolated, equilibrium galaxy models. In
turn, such gas density enhancement corresponds, in our modelling of
black hole accretion, to an increase in the MBH mass ratio. Applying
the standard dynamical friction formulation \citep{chandra43} to a
``naked'' MBH in our merger remnants, we estimate the timescale for
bringing the MBH down to the nuclear region ($<30$~pc) to be up to a
few billion years for separations up to $\sim1$~kpc, and comparable to
or longer than a Hubble time for larger separations. Thanks to a
combination of stellar-dynamical \citep[e.g.,][]{milomerritt01} and
gas-dynamical \citep[e.g.,][]{gould00,escala05,dotti07} processes, a
satellite MBH that sinks down to our current force resolution limit
may decay further and form a Keplerian binary together with the
primary MBH. Therefore, our results suggest that mergers that may most
promptly produce MBH binaries are also those that tend to enhance the
MBHs $q$, with respect to what inferred from the galaxy mass ratio
$q_{\rm gal}$.

Such increase in MBH $q$ happens in two distinct phases, whose
occurrence and relative importance depend on the details of the merger
process itself. Specifically, in the initial stages of the encounter,
the stronger tidal perturbations experienced by the satellite galaxy,
compared to those of the primary, cause an enhanced mass growth of its
MBH. In addition, in the last stages of the encounter, the orbit of
the secondary MBH may circularize inside the disk of the primary. As a
result, its ability to accrete gas and grow in mass relative to that
of the primary MBH can be further amplified. Such circularization and
associated increase in the accretion rate has been previously reported
in small-scale simulations of MBH pairs embedded in a common nuclear
disk \citep{dotti09a}. We note that the amount of gas left around the
satellite MBH by ram pressure stripping might be underestimated. In
fact, our simulations do not resolve the accretion disk around the
MBHs, nor the cold molecular phase in the nuclear region of the
satellite; both of these would be less susceptible to ram pressure
stripping \citep[e.g.,][]{quilis00} and could remain bound to the
satellite while the rest of its ISM is lost. In this case, the growth
of the secondary MBH may not stop completely and the mass ratios may
become even larger. 

A cautionary remark concerns the fact that MBH accretion has been
modelled assuming the Bondi-Hoyle-Lyttleton sub-grid accretion recipe
that is widely used in the literature. However, accretion will most
likely occur by means of angular momentum transport in a nuclear disk
with an effective $\alpha$-viscosity \citep[e.g.,][]{lin87}. Other
models for gas funnelling and accretion onto central MBHs have been
recently proposed to account for the shortcomings of the ``standard''
approach. \citet{power10} propose an accretion scheme for angular
momentum-dominated gaseous flows in a stationary gravitational
background, introducing a delay in the accretion as a free parameter
related to viscous timescales; from their simulations, they find that
the Bondi prescription \emph{overestimates} the inflow rate across the
MBH influence radius. \citet{hopkins10} model the transport of angular
momentum via disk instabilities arising in galaxy mergers, on scales
both larger and smaller than those resolved in this work. They
describe the inflow rates down to the MBH sphere of influence with
$\dot M\sim a M_{\rm g} \Omega$, where $M_{\rm g}$ is the available
gaseous mass, $\Omega$ the angular speed, and $a$ is related to the
amplitude of the instabilities. They find that the Bondi accretion
recipe systematically \emph{underestimates} (although with large
scatter) the actual inflow rate at sub-pc scales, while better
predictions are given by an effective $\alpha$-disk model. It is
therefore unclear whether our accretion recipe introduces a bias in
the MBH mass growth, and in which direction. In an attempt to
investigate the effect of other sub-grid recipes on our results, we
computed \emph{a posteriori} accretion rates in our reference
simulation with the recent $\alpha$-disk sub-resolution model of
\citet{debuhr09}. We find that such model yields higher $\dot M$
compared to the Bondi prescription, in agreement with the findings by
\citet{hopkins10}; moreover, the accretion rates onto the two MBHs are
enhanced by roughly the same factor. While the absolute values of
$\dot M$ can depend on the employed numerical prescription, the
physical picture concerning the \emph{relative} growth of the two MBHs
emerging from our simulations may not be substantially affected by the
specific choice of sub-grid modelling. Indeed, our findings reflect
clear and well-resolved large scale effects, namely how gravitational
torques and orbit circularization can enhance MBH sinking while making
a larger gaseous mass relative to $M_{\rm BH}$ available for accretion
onto the secondary black hole.

The results presented in this paper are especially relevant in the
context of MBH gravitational recoils \citep{lousto09,
tanaka09}. Indeed, if a large fraction of unequal-mass galaxy mergers
results in mergers between MBHs with nearly equal masses, then the
recoil velocity distribution of the MBH population will be different
than expected \citep[e.g.,][]{volonteri10}. However, the actual recoil
velocity distribution will also depend on the magnitude and relative
orientation of the spins of the MBHs at the final stage of the merger,
which is likely driven by gas dynamics at scales well below those
resolved in our simulations \citep{perego09,dotti10}.

Our findings, together with those of C09 and of \citet{kazantzidis05},
confirm the fundamental role of tidal stripping and gas-dynamical
effect in deciding the formation and properties of close MBH pairs in
unequal-mass galaxy mergers. Moreover, our results suggest that the
efficiency of MBH pair formation may correlate with the final mass
ratio of the pair itself, so that MBH pairs with larger mass ratios
tend to be produced more effectively and promptly. Gravitational wave
detectors, such as LISA, will enable the use of gravitational wave
signals from MBH coalescences as a new, independent probe of cosmic
structure formation. Indeed, gravitational wave-forms can allow the
determination of mass, spin, and orbital parameters of the merging
MBHs \citep{vecchio04}. In principle, this information could be used
to infer the masses of the merging host galaxies. However, our results
show that this connection cannot be made by simply applying the
observed scaling relations between the masses of MBHs and the
properties of their host galaxies, even if these scalings are
applicable to galaxies throughout cosmic history. The findings
presented here suggest that the mapping between galaxy and MBH mergers
depends on various factors, such as the gas content of the merging
galaxies and the encounter geometry, and as such might need to be
approached in a probabilistic way. Such investigations would require a
combination of a series of merger experiments that explore a larger
parameter space with semi-analytical models of the co-evolution
between galaxies and MBHs.

\acknowledgments{The authors are grateful to Jackson DeBuhr, Massimo
  Dotti, Marta Volonteri, and David Weinberg for stimulating
  discussions, and to the anonymous referee for comments that greatly
  improved this Paper.  SC, SK, and LM acknowledge the Kavli Institute
  for Theoretical Physics at the University of California at Santa
  Barbara for hosting the workshop ``Building the Milky Way'' during
  the initial stages of this work.  SC is also grateful to the Center
  for Cosmology and Astro-Particle Physics (CCAPP) at The Ohio State
  University for hospitality while completing this work. This research
  is supported by the Swiss National Science Foundation, by CCAPP, and
  by an allocation of computing time from the Ohio Supercomputer
  Center (OSC). Simulations were performed on the IBM Opteron Cluster
  ``Glenn'' at OSC and on the zBox3 at the University of
  Z\"urich. This research made use of the NASA Astrophysics Data
  System.}

\end{document}